# Simulation of pumping mechanism of $H_2O$-masers in circumstellar envelopes of late-type stars

**A V Nesterenok**

Ioffe Physical-Technical Institute, 26 Polytechnicheskaya St., 194021, Saint Petersburg, Russia

St. Petersburg State Polytechnic University, 29 Polytechnicheskaya St., 195251, Saint Petersburg, Russia

E-mail: alex-n10@yandex.ru

**Abstract.** The radiative transfer problem is considered in dense gas-dust clouds. Physical parameters are adopted in simulations corresponding to gas-dust clouds in the circumstellar envelopes of late-type stars. An one-dimensional plane-parallel slab geometry of the cloud is considered. The radiative transfer problem is solved using the accelerated lambda iteration technique. The rotational levels of the five lowest vibrational levels of $H_2O$-molecule are considered in the simulations. The level populations of $H_2O$-molecule are computed as a function of distance in the cloud. The conditions of the level population inversion occurrence are investigated. In particular, the factors controlling the population inversion in the 22.2 GHz maser line are considered. A radiative excitation of the molecules by external radiation field of a star is considered together with the collisional pumping of the maser.

## 1. Introduction

Water maser emission has been detected in many astrophysical objects – active galactic nuclei, star formation regions, circumstellar envelopes of late-type stars. In this paper we will focus on the water masers in the envelopes of the late-type stars.

Stars with masses up to a few solar masses evolve onto the Asymptotic Giant Branch after the hydrogen fuel has been exhausted in its interior. The stars become very large (> 1 AU) and cool (≈ 2500 K). Stars with higher mass become Red Supergiants and have ten times the size of AGB stars. The evolution of late-type stars is dominated by the mass loss. The photosphere and expanding circumstellar envelope are effective producers of a variety of molecular species. Oxygen-rich stars produce SiO, $H_2O$ and OH masers with spatially very compact, spectrally narrow lines. The high intensity and the narrowness of spectral features of a maser allow one to obtain detailed maps of the distribution of maser sources. The $H_2O$ masers around late-type stars are found at the distances of about 5-50 stellar radius from the central star [1]. They provide a very high resolution probe of the structure, kinematics and conditions in the circumstellar envelope.

In this study a simulation is done of the pumping process of water masers in a circumstellar envelope of late-type star. Two mechanisms of molecule excitation are considered – collisional excitation of the molecules and excitation by the radiation of the central star.

## 2. The method

*2.1. Radiative transfer and population balance*

Each energy level $i$ of the molecule is characterized by its statistical weight $g_i$, energy $\varepsilon_i$ and the population $n_i$. The energy relations $\varepsilon_i > \varepsilon_j$ and $\varepsilon_i < \varepsilon_j$ are abbreviated by $i > j$ and $i < j$, respectively. The equations of statistical equilibrium for the level populations may be written

$$n_i \sum_{j \neq i}^{M} \left( R_{ij} + C_{ij} \right) - \sum_{j \neq i}^{M} n_j \left( R_{ji} + C_{ji} \right) = 0 \qquad (1)$$

where $C_{ij}$ are the collisional rate coefficients and $R_{ij}$ represent radiative rate coefficients, $M$ is the total number of energy levels involved in the model. The values of $R_{ij}$ are given by $R_{ij} = A_{ij} + B_{ij} \bar{J}_{ij}$ for $i > j$ and $R_{ij} = B_{ij} \bar{J}_{ij}$ for $i < j$, $A_{ij}$ and $B_{ij}$ are the Einstein coefficients, $\bar{J}_{ij}$ is the line-weighted mean intensity, defined by

$$\bar{J}_{ij} = \frac{1}{4\pi} \int_0^{4\pi} d\Omega \int_0^{\infty} d\nu \, \varphi(\Omega,\nu) I(\Omega,\nu)$$

where $I(\Omega,\nu)$ is the intensity at frequency $\nu$ and angle $\Omega$, $\varphi(\Omega,\nu)$ is the normalized line profile function. In the plane parallel geometry the transfer equation for the intensity is

$$\mu \frac{\partial I(\mu,\nu)}{\partial z} = -\chi(\mu,\nu) I(\mu,\nu) + \eta(\mu,\nu) \qquad (2)$$

where $z$ is the coordinate along the normal to the cloud plane, $\mu = 1/\cos\theta$, $\theta$ is the angle between the normal to the cloud plane and a given direction, $\chi(\mu,\nu)$ and $\eta(\mu,\nu)$ are the total opacity and emissivity at frequency $\nu$ and direction $\mu$. The emissivity and opacity in the neighbourhood of a given line $i \rightarrow j$ depend on the optical properties of the dust and level populations $n_i$ and $n_j$ [2]. The problem of calculating the populations of energy levels consists in solving the set of rate equations (1) coupled to the equations of radiative transfer (2).

Various techniques have been developed to yield a self-consistent set of level populations and radiation fields [3]. We have adopted the accelerated lambda iteration (ALI) method for molecular level calculation. The detailed description of the method was given by [2]. In brief outline, the equations are solved iteratively starting with any initial level populations. The mean intensities computed with the aid of equation (2) are substituted to the statistical equilibrium equations (1) and new level populations are calculated. The procedure is repeated until the relative change in the level populations or mean intensities between two successive steps falls below a given convergence criterion. The concept of the lambda operator is introduced. The lambda operator represents the entire procedure of computing the line-weighted mean intensities $\bar{J}_{ij}$ from the source function

$$\bar{J}_{ij} = \Lambda_{ij} \left[ S^{\dagger}(\Omega,\nu) \right]$$

where $S^{\dagger}(\Omega,\nu)$ is the source function from the previous iteration step. The acceleration of iteration scheme is achieved by splitting the lambda operator onto the local contribution to the intensity and the remainder [2]. The calculation of the line-weighted mean intensity at each iteration step is modified

$$\bar{J}_{ij} = \left( \Lambda_{ij} - \Lambda_{ij}^* \right) \left[ S^{\dagger}(\Omega,\nu) \right] + \Lambda_{ij}^* \left[ S(\Omega,\nu) \right] \qquad (3)$$

This is inserted into the equations of statistical equilibrium (1) and the equations are solved for the new level populations. Note, that the equation (3) contains the new populations in the last term which have to be found in the end of iteration step. In addition to this operator splitting technique, one can apply iteration improvement schemes such as the Ng acceleration [2].

*2.2. Collisional rate coefficients and spectroscopic data*

We calculate the populations for the 350 rotational levels of the five lowest vibrational levels of $H_2O$ molecule. The energy of the uppermost level is about 4800 cm$^{-1}$, or 6900 K in units of temperature. The spectroscopic data for the $H_2O$ molecule is taken from the HITRAN 2008 database [4]. The rate coefficients for $H_2O$ collisional transitions in inelastic collisions of $H_2O$ with $H_2$ were taken from [5]. The rate coefficients for $H_2O$ collisional transitions in collisions of $H_2O$ with He were taken from [6]. The data [6] are restricted to 45 lower $H_2O$ levels. For higher $H_2O$ levels collisional rate coefficients are taken to be scaled rate coefficients $H_2O$-$H_2$ with the scaling factor being 0.74. The factor accounts for the different reduced mass when He is the collisional partner. The helium-to-hydrogen ratio was assumed to be He/H = 0.1. The rate constants of inelastic collisions involving $H_2O$ molecule and atomic hydrogen H are assumed to be a factor 1.15 larger than He-$H_2O$ rate coefficients. This value takes both the smaller mass and the smaller cross section of H atom into account.

*2.3. Physical model*

The physical parameters adopted in the calculations are representative for the gas clumps in the wind of AGB stars [1]. The distance between the masing cloud and the star is taken to be $D = 10 R_S$ where $R_S = 4 \times 10^{13}$ cm – the star radius. An one-dimensional plane-parallel slab geometry of the cloud is considered. The cloud thickness $H$ is chosen to be $5 \times 10^{13}$ cm. The cloud is divided into $N = 100$ layers, using a distance scale defined by

$$z_{k+1} - z_k = a(z_k - z_{k-1}), \quad z_{k+1} - z_k = z_{N-k} - z_{N-k-1}, \quad k = 1,\ldots,N/2-1,$$

where $k$ is the depth point number, $a$ is the coefficient ($a > 1$). For the boundary layers the size is chosen such that the maximum optical depth in any line is low, $z_1 - z_0 = z_N - z_{N-1} = H/10^6$.

The mass loss rate of the star is adopted to be $\wp = 3 \times 10^{-6}$ of solar masses per year. One may estimate the gas density of the outflow based on the mass conservation equation

$$\rho = \wp \left(4\pi r^2 \upsilon\right)^{-1}$$

where $\rho$ is the gas density, $r$ is the distance from the star, $\upsilon$ is the outflow velocity. Taking the gas outflow velocity of $\upsilon = 10$ km/s, for the hydrogen atom concentration (both unbound and bound in molecular hydrogen) we have $n_H = 4 \times 10^7$ cm$^{-3}$ at the distance in question. It's suggested that the densities of $H_2O$ maser clouds are higher than the average wind densities [1]. We have used three values of the total hydrogen atom concentration in the maser cloud – $4 \times 10^7$ cm$^{-3}$, $1.2 \times 10^8$ cm$^{-3}$ and $4 \times 10^8$ cm$^{-3}$. The ratio of concentrations of molecular and atom hydrogen is adopted to be 1. The $H_2O$ abundance is taken to be equal to the average solar system abundance of oxygen – $5.4 \times 10^{-4}$ [7], the ortho- to para-$H_2O$ ratio is 3. The velocity of turbulent motions of the gas is adopted to be 1 km/s, velocity gradient in the cloud – $10^{-9}$ s$^{-1}$ [1].

We consider the fixed size of the dust grains 0.1 μm. We use the complex dielectric function for the dust material presented in [8]. The absorption efficiency is computed from Mie theory for spherical grains (program BHMIE from [9]). The dust to gas mass ratio is taken to be 0.01.

The gas and dust temperatures in the cloud depend on cooling and heating processes such as gas-dust collisions, radiative cooling and heating by starlight. We use the calculation results from [10,11]. The dust and gas temperatures are taken to be equal 400 K and 600 K, respectively.

The stellar temperature is taken to be 2500 K. The radiation field intensity of the star is assumed to have a blackbody spectrum. The stellar radiation is taken into account in the boundary conditions for the radiation transfer differential equation (2).

Using the ALI method with a local operator and Ng acceleration, the system is iterated towards a solution with a convergence criterion of $10^{-4}$ in relative change in level populations for successive iterations.

### 3. Calculation results

The $H_2O$ maser line at 22.2 GHz corresponds to the transition between rotational levels of the ground vibrational state of ortho-$H_2O$ molecule $6_{16} - 5_{23}$. The Figure 1 presents the calculated gain at the line profile center for the maser line 22.2 GHz. The spectral gain profile broadening due to the hyperfine splitting is taken into account [12]. The average values of the gain are $1.1\times10^{-13}$, $0.9\times10^{-13}$ and $0.8\times10^{-13}$ cm$^{-1}$ for the values of total hydrogen atom concentration $4\times10^7$, $1.2\times10^8$ and $4\times10^8$ cm$^{-3}$, respectively. The average value of the gain decreases with increasing gas density.

The effect of the stellar radiation field on the maser pumping process is studied. The figure 2 presents the calculation results for the gain in two cases – with and without taking into account the stellar radiation in the radiative transfer. The total hydrogen atom concentration is taken to be $1.2\times10^8$ cm$^{-3}$. The average values of the gain are $0.9\times10^{-13}$ and $1.3\times10^{-13}$ cm$^{-1}$, with the gain being lower when the stellar radiation is taken into account. The result contradicts to the generally accepted conclusion that the stellar radiation has a minimal effect on the pumping process of the 22 GHz maser [13].

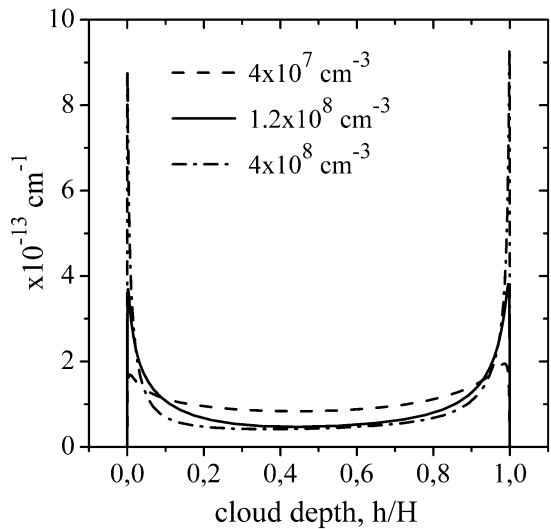
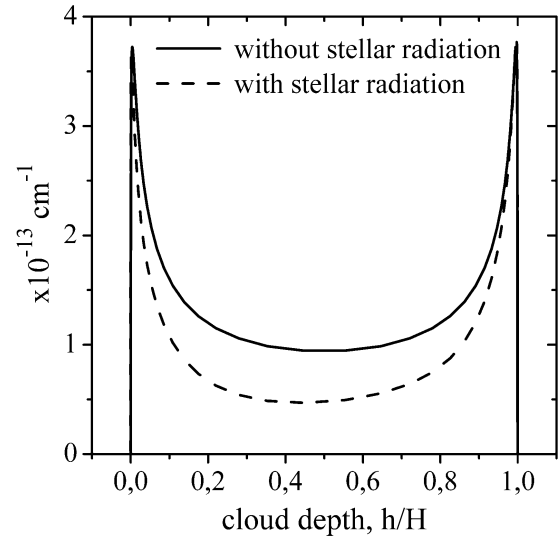

**Figure 1.** The gain in the 22 GHz maser line as a function of the cloud depth. The results for three values of total hydrogen atom concentration are presented.

**Figure 2.** The gain in the 22 GHz maser line as a function of the cloud depth. The results are given for the calculations with and without taking into account stellar radiation in the radiative transfer.

### 4. Conclusions

In the present work, accelerated lambda iteration method has been used to calculate the level populations of $H_2O$ molecule in the dense gas-dust clouds. The physical conditions encompassing in the inner part of circumstellar envelopes are adopted in the simulations. A radiative excitation of the molecules by radiation field of a star is considered in the model. The gain in the maser line of ortho-$H_2O$ molecule 22.2 GHz is calculated. It is found that the effect of the stellar radiation on the gain in the maser line is substantial. The radiation field of the star has to be taken into account in the simulations of the maser pumping process.

**Acknowledgments**
This work was supported by the Russian Foundation for Basic Research (project no. 11-02-01018a), the Program of the President of Russia for Support of Leading Scientific Schools (project no. NSh-4035.2012.2), by Ministry of Education and Science of Russian Federation - Agreement No.8409, 2012 and contract # 11.G34.31.0001 with SPbSPU and leading scientist G.G. Pavlov.